\documentclass[12pt,preprint]{emulateapj}
\usepackage{graphicx}

\shorttitle{Discovery of a Low-Mass Companion}
\shortauthors{Stephen R. Kane et al.}
\slugcomment{Submitted for publication in the Astrophysical Journal}
\newcommand{\nc}{\newcommand}
\nc{\teff}{$T_{\rm eff}$\,}
\nc{\logg}{log\,$g$\,}
\nc{\kms}{\,${\rm km\,s}^{-1}$\,}
\nc{\mic}{$\xi_{\rm t}$\,}

\begin{document}

\title{Discovery of a Low-Mass Companion to the Solar-Type Star TYC
  2534-698-1}

\author{Stephen R. Kane\altaffilmark{1}, Suvrath
  Mahadevan\altaffilmark{2}, William D. Cochran\altaffilmark{3},
  Rachel A. Street\altaffilmark{4}, Sivarani
  Thirupathi\altaffilmark{2}, Gregory W. Henry\altaffilmark{5},
  Michael H. Williamson\altaffilmark{5}}
\email{skane@ipac.caltech.edu}
\altaffiltext{1}{NASA Exoplanet Science Institute, Caltech, MS 100-22,
  770 South Wilson Avenue, Pasadena, CA 91125}
\altaffiltext{2}{Department of Astronomy, University of Florida, 211
  Bryant Space Science Center, Gainesville, FL 32611-2055}
\altaffiltext{3}{Department of Astronomy, University of Texas, Austin,
  TX 78712}
\altaffiltext{4}{Las Cumbres Observatory Global Telescope, Goleta, CA
  93117}
\altaffiltext{5}{Center of Excellence in Information System, Tennessee
  State University, 3500 John A. Merritt Blvd., Box 9501, Nashville,
  TN 37209}


\begin{abstract}

Brown dwarfs and low-mass stellar companions are interesting objects
to study since they occupy the mass region between deuterium and
hydrogen burning. We report here the serendipitous discovery of a
low-mass companion in an eccentric orbit around a solar-type main
sequence star. The stellar primary, TYC 2534-698-1, is a G2V star
that was monitored both spectroscopically and photometrically over
the course of several months. Radial velocity observations indicate a
minimum mass of 0.037 $M_\odot$ and an orbital period of $\sim 103$
days for the companion. Photometry outside of the transit window shows
the star to be stable to within $\sim 6$ millimags. The semi-major
axis of the orbit places the companion in the 'brown dwarf desert' and
we discuss potential follow-up observations that could constrain the
mass of the companion.

\end{abstract}



\section{Introduction}

Brown dwarfs in short and intermediate periods ($<5$ AU) are
relatively rare compared to their occurrence at wider
separations. \citet{met08} show that the frequency of brown dwarfs at
larger separations (29-1590 AU) is $\sim$ 3.2\%, much larger than the
$\sim 0.5$\% observed for separations less than 3 AU
\citep{mar00}. This 'brown dwarf desert' in radial velocity
observations is highly significant and not easy to
explain. Confirmation of radial velocity detected brown dwarf
candidates is also challenging since the observations only yield a
minimum mass and radial velocity alone cannot break the degeneracy
between companion mass and inclination.

For some of the brighter stars hosting brown dwarf candidates, the
Hippparcos measurements \citep{per97} (coupled with the radial
velocity data) can be used to estimate the mass of the companion. Such
techniques have been demonstrated by \citet{zuc00}, \citet{ref06}, and
\citet{kur08}. Eclipsing binary systems, and close binaries where both
spectra can be observed, offer the prospect of the dynamical
measurement of the mass of the two companions, and such systems have
been used to determine accurate brown dwarf and low-mass star masses
by \citet{zap04} and \citet{sta06}. \citet{ire08} have measured the
dynamical mass for GJ~802b using aperture masking and interferometry.

Another way to confirm the mass of the companion unambiguously is to
detect a transit. Coupled with radial velocity data this constrains
the inclination angle and therefore the mass of the companion. The
challenge with attempting this method on brown dwarf candidates
discovered using the radial velocity technique is that the periods are
typically 100--1000 days, making the transit probability very
small. Another approach is to follow-up up detections from transit
surveys with radial velocity observations to determine the mass of the
candidate. This method has recently been employed by \citet{del08} in
the discovery of the transiting brown dwarf ($\sim$22 Jupiter mass)
companion to the F3V star CoRoT-Exo-3.  While this is frequently done
for the most promising candidates, the large amount of time required
for precision radial velocity observations to confirm a planet leads
to many candidates also being rejected before radial velocity
follow-up. We selected 5 candidate transiting objects from the
SuperWASP fields between 06--16 hours \citep{kan08} for radial
velocity follow-up. Our motivation was to confirm either a brown dwarf
or a low-mass stellar companion. No confirmed transiting brown dwarf
around relatively bright stars ($V< 12$) has been discovered yet,
although the planet XO-3b \citep{joh08} may be massive enough for
deuterium burning. Transiting low-mass stars are also interesting in
their own right since an estimate of both mass and radii helps to
constrain the equation of state used in current models
\citep[e.g.,][]{cha00}. In this paper we present our radial velocity
and photometric observations for one of our candidates, TYC
2534-698-1, which has a Tycho $V$ magnitude of 10.8. For this object
the SuperWASP data shows a transit depth of 17.9 millimags and a
periodicity of $\sim 2.67$ days. We find no radial velocity
variability at this period, but our radial velocities show the
existence of a 0.037 $M_\odot$ low-mass companion (minimum mass) at a
103.69 day orbit.


\section{Observations}

In this section we discuss the acquisition of the radial velocity and
photometric data.


\subsection{Radial Velocity}

Radial velocity observation of TYC 2534-698-1 were obtained using the
High Resolution Spectrograph \citep[HRS,][]{tul98} on the Hobby-Eberly
Telescope (HET) in queue scheduled mode. The observations used a 2
arcsecond fiber, resulting in a spectral resolution of
$R = 60,000$. Thirteen observations were acquired with the HRS with an
iodine cell placed in the stellar beam path. One stellar template
exposure was acquired without the iodine cell in order to facilitate
the extraction of precise radial velocities from the combined star and
iodine spectra. On sky integration time ranged from 10--15 minutes.
The queue scheduled nature of the HET allows great flexibility in the
observations. The observations were placed on hold many times while
the data were reduced in order to determine the best period. This made
it possible to request observations that would be most useful in
constraining the period of the companion. Our initial cadence was
designed to detect the signature of the claimed SuperWASP periodicity
of $P \sim 2.67$ days. The high-resolution spectroscopic data obtained
with the iodine cell were reduced by the method described in
\citet{coc04} using the stellar template and high-resolution iodine
spectrum to simultaneously model instrumental and doppler velocity
shifts as well as variations in the point-spread function of the
spectrograph. This technique allows the instrument drift to be
calibrated out, and radial velocity precision of better than 2--3
ms$^{-1}$ has been demonstrated on bright stars. Good velocity
precision can only be obtained in the spectral region where iodine has
sharp absorption lines, and therefore only the 5000--6200\AA\ region
is used to derive velocities. Table \ref{tbl:vels} lists the radial
velocities obtained for TYC 2534-698-1. The dominant source of
velocity error is the photon noise limited radial velocity
uncertainty, mainly due to the target being faint and the exposure
time being short. The error bars stated are determined from the the
scatter in the velocities obtained from 2\AA\ chunks of the spectrum.

The radial velocities are not consistent with the expected period from
the SuperWASP data, but show the presence of a longer period
companion. We discuss the fits to the radial velocity data in \S
\ref{sec:rvmodel}.

\begin{table}
  \begin{center}
    \caption{HET Radial Velocities for TYC 2534-698-1}
    \label{tbl:vels}
    \begin{tabular}{@{}ccc}
      \hline
      Julian Date & Velocity (ms$^{-1}$) & Velocity Error (ms$^{-1}$) \\
      \hline
      2454515.809038 & -2449.63 & 29.33 \\
      2454516.802895 & -2381.23 & 21.02 \\
      2454518.789397 & -1955.85 & 21.47 \\
      2454523.797391 &  -705.29 & 22.06 \\
      2454550.939348 &   227.73 & 22.87 \\
      2454551.938358 &   185.48 & 23.33 \\
      2454552.929063 &   149.75 & 23.74 \\
      2454553.707716 &   133.50 & 23.35 \\
      2454554.699371 &    99.05 & 21.60 \\
      2454562.909758 &  -271.73 & 24.55 \\
      2454575.875428 &  -853.48 & 23.58 \\
      2454611.773732 & -2958.58 & 25.17 \\
      2454627.726562 &  -663.11 & 28.62 \\
      \hline
    \end{tabular}
  \end{center}
\end{table}


\subsection{Photometry}
\label{sec:phot}

Our target was monitored photometrically for over 30 days with three
separate telescopes. The objectives of these observations were
twofold: firstly to provide confirmation of the transit of the
low-mass companion, and secondly to determine the photometric
stablility of the target. The telescopes used were an automated
14-inch imaging telescope mounted at Dyer Observatory, Tennessee, a
16-inch telescope located at Las Cumbres Observatory (LCO) in
California, and the Tenagra II automated 32-inch telescope located in
Arizona.

As radial velocity data were obtained, the predicted time of transit
was calculated based on the current best fit to the data (see \S
\ref{sec:ephem}). Since the best-fit orbital period shifted
substantially from the original SuperWASP period, the time of
predicted transit changed several times over the course of the HET
observations. As such, photometry was obtained over a series of nights
as the model parameters for the orbit were updated.

\begin{figure}
  \includegraphics[angle=270,width=8.2cm]{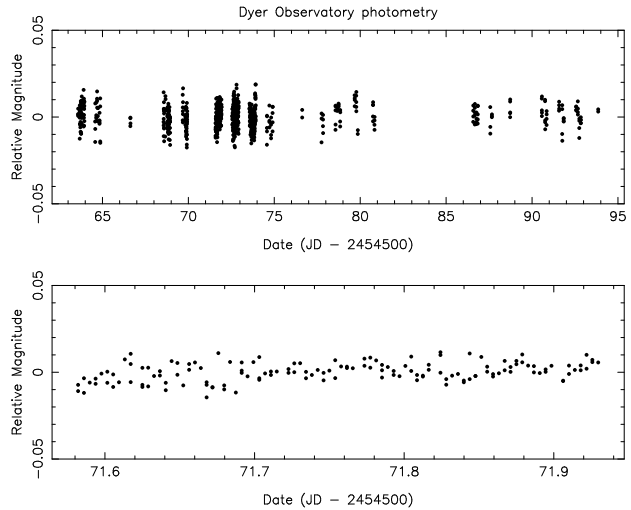}
  \caption{Photometry of TYC 2534-698-1 obtained using an automated
    14-inch imaging telescope at Dyer Observatory.}
  \label{phot_dyer}
\end{figure}

Shown in Figure \ref{phot_dyer} are the photometric measurements
obtained using the Dyer Observatory 14-inch telescope coupled to an
SBIG ST-1001E CCD camera. The top panel presents the photometry
acquired over the course of $\sim 30$ days, and the bottom panel
presents the photometry from a single night. The relative magnitudes
were produced by comparing the brightness of TYC 2534-698-1 to the
mean brightness of 6 constant stars in the same field. The star
appears to be quite stable, with a standard deviation of 6.2
millimags. No evidence of a transit was detected in this dataset.

\begin{figure}
  \includegraphics[angle=270,width=8.2cm]{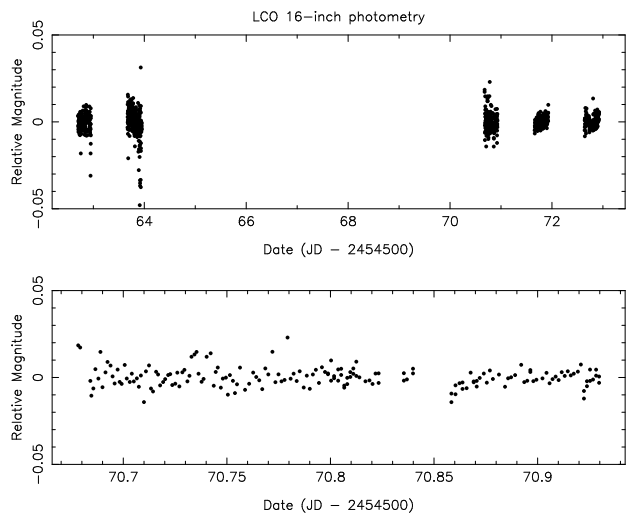}
  \caption{Photometry of TYC 2534-698-1 obtained using a 16-inch
    telescope mounted at Las Cumbres Observatory.}
  \label{phot_lco}
\end{figure}

Figure \ref{phot_lco} shows the photometry obtained with the LCO
16-inch telescope covering $\sim 10$ days. Once again, the star is
very stable on the dates observed, with a standard deviation of 5.7
millimags. No transit signature was detected on any of the nights
observed.

\begin{figure}
  \includegraphics[angle=270,width=8.2cm]{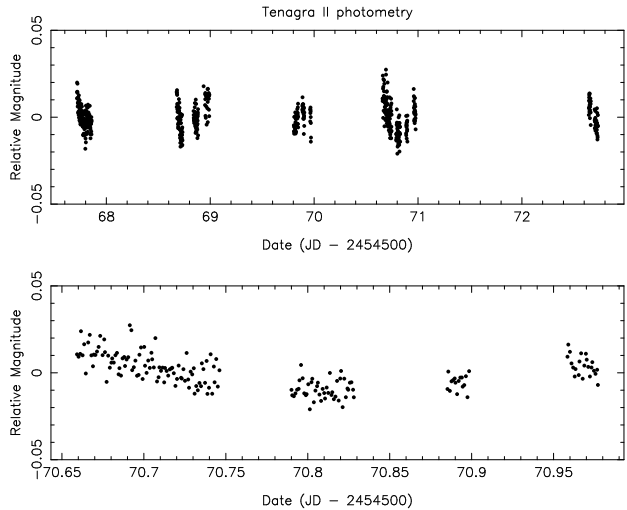}
  \caption{Photometry of TYC 2534-698-1 obtained using the Tenagra II
    32 inch telescope.}
  \label{phot_ten}
\end{figure}

Figure \ref{phot_ten} presents the photometry from the Tenagra II
telescope, covering a period of $\sim 6$ days shown in the top panel,
and from a single night shown in the bottom panel. The relative
magnitudes were computed using a single comparison star due to the
limited field-of-view. These data also do not detect any transit
event, but confirm the photometrically stable nature of the star. In
this case, the standard deviation is 7.9 millimags, with residual
airmass effects slightly increasing the nightly scatter.

\section{Data Analysis}

In this section we present the results of analysing the spectra,
radial velocity and photometric data, deriving the stellar parameters,
best-fit orbital solution, mass, and transit ephemeris.


\subsection{Stellar Parameters \& Mass Estimate}

While TYC 2534-698-1 has no Hipparcos parallax estimate, it does have
TYCHO-2 photometry ($B_T, V_T$) as well as 2MASS photometry ($J$, $H$,
$K$). TYCHO-2 photometry is known to degrade beyond $V_T > 11.0$, but
our target is bright enough that the photometric errors are not too
large. Table \ref{tab:photometry} lists the photometric magnitudes and
associated errors in the 5 different filters considered. To derive an
approximate stellar temperature ($T_{\rm eff}$) we fit a series of
synthetic SEDs to the photometric data using the CHORIZOS code
\citep{mai04}. This yields a best-fit temperature of 5700 K.

\begin{figure}
  \begin{center}
    \includegraphics[angle=0,width=3.5in]{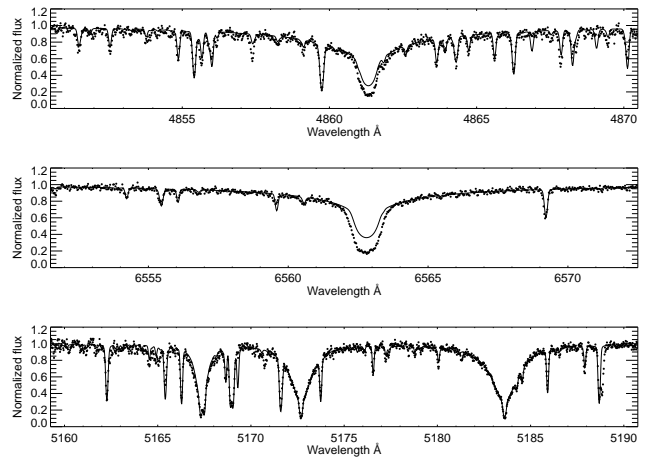}
    \caption{Continuum normalized plot of the high-resolution spectrum
      obtained with the HRS instrument. Wavelength regions shown are
      centered on H$\alpha$, H$\beta$, and Mg b. The best-fit stellar
      model (see Table \ref{tab:stellarparams}) is plotted as a solid
      line.}
    \label{fig:stellarmodel}
  \end{center}
\end{figure}

A more precise estimation of the spectral parameters was then
performed using the stellar template obtained as part of the radial
velocity observing program. The template was taken without the iodine
cell in the beam path and allows the use of the entire spectral range
covered by the HRS. We use the latest MARCS model atmospheres
\citep{gus08} for the analysis. Generation of synthetic spectra and
the line analysis were performed using the turbospectrum code
\citep{alv98}, which employs line broadening according to the
prescription of \citet{bar98}. The line lists used are drawn from a
variety of sources. Updated atomic lines are taken mainly from the
VALD database \citep{kup99}. The molecular species CH, CN, OH, CaH and
TiO are provided by B. Plez \citep[see][]{ple05}, while the NH, MgH
and ${\bf\rm C_2}$ molecules are from the Kurucz
linelists\footnote{http://kurucz.harvard.edu/LINELISTS/LINESMOL/}.
The solar abundances used here are the same as \citet{asp05}. We use
FeI excitation equilbrium and derived an effective temperature $T_{\rm
  eff} = 5700\pm80$K which is consistent with the effective
temperature derived from photometry. We also fit the H$\alpha$ and
H$\beta$ line wings, and derived $T_{\rm eff} = 5700\pm 100$K, which
is also consistent with the $T_{\rm eff}$ derived from photometry and
SED. We also find \logg = $4.5 \pm 0.1$, based on ionization
equilibrium of FeI and FeII lines and by fitting the wings of the Mgb
lines. A microturbulence value \mic= 1.6\kms is derived by forcing
weak and strong FeI lines to give the same abundances. We only used
the FeI lines weaker the 120m\AA\ in the analysis. These parameters
are consistent with a G2V spectral type with a relatively low
metallicity. To estimate the mass of the star we used Padova
isochrones for metallicities of $Z = 0.008, 0.019, 0.030$ and ages
between 1.778 and 7.943 Gyrs. Taking into account the uncertainities
in \teff and \logg, from these isochrones we derive a mass estimate of
$0.998 \pm 0.040 M_{\odot}$. Figure \ref{fig:stellarmodel} shows the
observed high-resolution spectrum, as well as the best-fit model, for
wavelength regions centered on H$\alpha$, H$\beta$, and Mg~b. Table
\ref{tab:stellarparams} lists the derived stellar parameters.

\begin{table}
  \begin{center}
    \caption{TYCHO-2 \& 2MASS Photometry for TYC 2534-698-1}
    \label{tab:photometry}
    \begin{tabular}{@{}lll}
      \hline
      Filter & Magnitude & Source \\
      \hline
      $B_T$ & $11.567\pm0.061$ & TYCHO-2 \\
      $V_T$ & $10.779\pm0.051$ & TYCHO-2 \\
      $J$ & $9.501\pm0.021$  & 2MASS \\
      $H$ & $9.226\pm0.022$  & 2MASS \\
      $K$ & $9.127\pm0.019$  & 2MASS \\
      \hline
    \end{tabular}
  \end{center}
\end{table}

\begin{table}
  \begin{center}
    \caption{Stellar Parameters}
    \label{tab:stellarparams}
    \begin{tabular}{@{}lc}
      \hline
      Parameter & Value \\
      \hline
      \teff     & $5700 \pm 80$K \\
      \logg     & $4.5 \pm 0.1$ \\
      \mic      & $1.6$ km s$^{-1}$ \\
      $M_\star$ & $0.998 \pm 0.040 M_{\odot}$ \\
      $[Fe/H]$  & $-0.25 \pm 0.06$\\
      \hline
    \end{tabular}
  \end{center}
\end{table}


\subsection{Model Fitting}
\label{sec:rvmodel}

The radial velocity data were fit using the iterative grid-search
approach described by \citet{kan07}. Fortunately, the allocated HET
time allowed us to obtain complete phase coverage of the orbit, thus
removing ambiguity at longer periods. As more measurements were
obtained, the eccentric nature of the orbit quickly became clear,
which necessitated even greater phase coverage in order to constrain
the shape of the radial velocity variation.

\begin{figure}
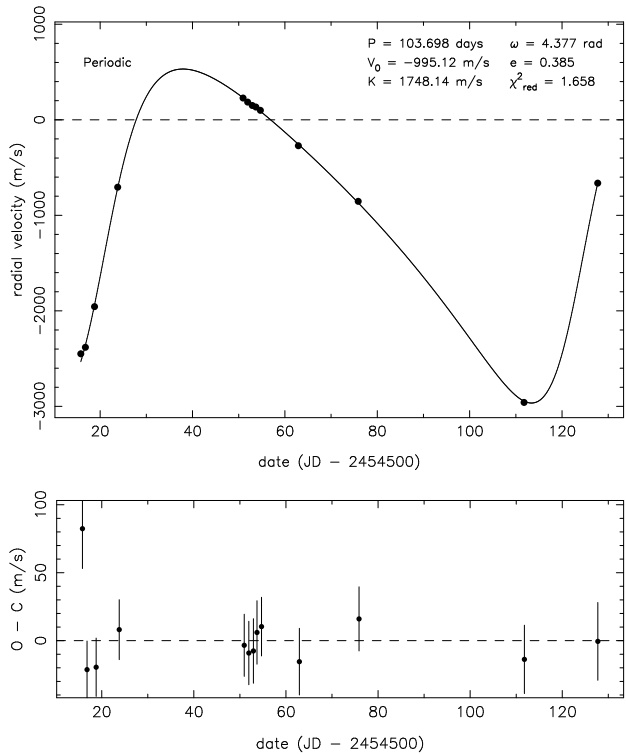

  \begin{center}
    \begin{tabular}{c}
      \includegraphics[angle=270,width=8.2cm]{f5a.eps} \\
      \includegraphics[angle=270,width=8.2cm]{f5b.eps}
    \end{tabular}
  \end{center}
  \caption{Radial velocity measurements of TYC 2534-698-1 along with
    the best-fit orbital solution (solid line). Error bars are shown
    but are small compared with the radial velocity amplitude in this
    case. The lower panel shows the residuals of the fit, observed
    minus calculated ($O - C$).}
  \label{fig:rvmodel}
\end{figure}

Shown in Figure \ref{fig:rvmodel} are the complete set of radial
velocity measurements obtained at the HET as part of this
program. Over-plotted is a solid line which indicates the best-fit
orbital model to the data, with a reduced $\chi^2$ of 1.658. The lower
panel shows the residuals from fitting the data with this model. The
orbital parameters that provide the best-fit solution are shown in
Table \ref{tab:rvmodel}. The higher than expected eccentricity caused
us to extend observations of this target until we were about to
observe the turn-around, which occurred at around JD 2454613. Using
simulated data points, we calculated the optimal times of observations
for the final few measurements which would yield the greatest
constraints on the orbital period.

\begin{table}
  \begin{center}
    \caption{The best-fit orbital parameters, including period $P$,
      systemic velocity $V_0$, semi-amplitude $K$, argument of
      periastron $\omega$, eccentricity $e$, and time at periastron
      $t_0$.}
    \label{tab:rvmodel}
    \begin{tabular}{@{}lc}
      \hline
      Parameter & Value \\
      \hline
      $P$      & $103.698 \pm 0.111$ days\\
      $V_0$    & $-995.12 \pm 18.03 \ \mathrm{m \ s^{-1}}$\\
      $K$      & $1748.14 \pm 27.48 \ \mathrm{m \ s^{-1}}$ \\
      $\omega$ & $4.377 \pm 0.013$ rad\\
      $e$      & $0.385 \pm 0.011$ \\
      $t_0$    & JD $2454519.389 \pm 0.137$ \\
      \hline
    \end{tabular}
  \end{center}
\end{table}

\begin{figure}
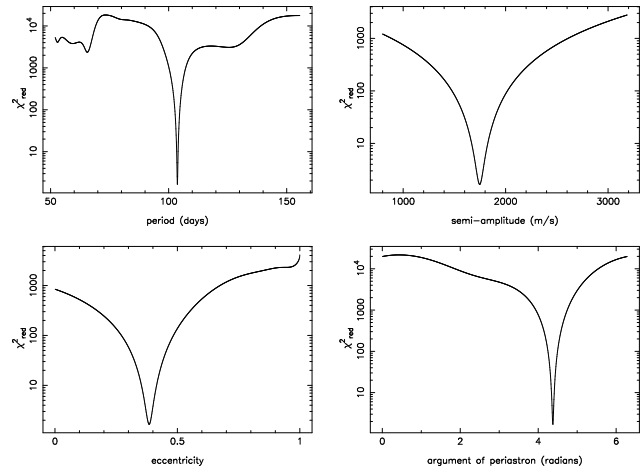

  \begin{center}
    \begin{tabular}{cc}
      \includegraphics[angle=270,width=4.0cm]{f6a.eps} &
      \includegraphics[angle=270,width=4.0cm]{f6b.eps} \\
      \includegraphics[angle=270,width=4.0cm]{f6c.eps} &
      \includegraphics[angle=270,width=4.0cm]{f6d.eps} \\
    \end{tabular}
  \end{center}
  \caption{$\chi^2$ maps of parameter space showing the locations of
    the local and global minima for the period, semi-amplitude,
    eccentricity, and argument of periastron.}
  \label{chi2maps}
\end{figure}

Given the phase coverage and eccentricity for this system, it is
important to determine the reliability of the fitted parameters. By
producing $\chi^2$ maps of parameter space, one can see where the
local and global minima lie in relation to the fit values for those
parameters. This is achieved by varying one parameter and holding all
of the others fixed. Figure \ref{chi2maps} shows the $\chi^2$ maps for
the period, semi-amplitude, eccentricity, and argument of periastron.
The phase coverage of the orbit produces significant global minima
which match the best-fit values shown in Table \ref{tab:rvmodel}. The
acquistion of the last two radial velocity measurements in particular
allowed us to constrain the value of the period.

\subsection{Companion Inclination and Mass Estimate}

The mass function can be related to the observed period, eccentricity,
and radial velocity semi-amplitude as:
\begin{equation}
  \frac{(m \sin{i})^3}{(M_\star+m)^2} =
  \frac{P(1-e^2)^{\frac{3}{2}}K^3}{2\pi G}
  \label{eqn:massfunction}
\end{equation}
where $M_\star$ is the mass of the primary and $m$ the mass of the
secondary. Without having observed the transit we cannot constrain
$\sin{i}$ directly, leading to the degeneracy of the companion mass
with assumed inclination angle. Using the derived stellar mass ($0.998
\pm 0.040 M_{\odot}$) for the primary with the orbital parameters
determined from the radial velocity model (Table \ref{tab:rvmodel}) we
determine that the minimum mass of the companion for an edge-on orbit
($\sin{i} = 1$) is $0.0373 \pm 0.011 M_{\odot}$. Using an upper-mass
limit of $0.08 M_\odot$ for a brown dwarf, Equation
\ref{eqn:massfunction} predicts that a companion of this mass would
have an orbital inclination of $i=28.5^{\circ}$ to reproduce the
radial velocity curve observed.

\begin{figure}
  \includegraphics[angle=0,width=8.2cm]{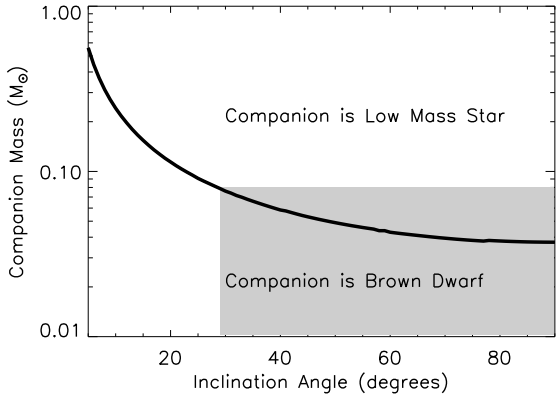}
  \caption{This figure shows the mass of the companion for different
    values of the inclination angle ($i$). The mass of the companion
    falls in the brown dwarf regime for all inclination angles larger
    than 28.5 degrees.}
  \label{fig:exclusion_regions}
\end{figure}

For a chance orientation of the orbital inclination, the probability
that the inclination angle $i$ is less than an angle $\theta$ is given
by
\begin{equation}
  p(i<\theta) = 1 - \cos{\theta}
\end{equation}
The probability of the companion having an inclination less than
$28.5^{\circ}$, and therefore being more massive than a brown dwarf,
is only 12.1\%. Figure \ref{fig:exclusion_regions} shows the range of
possible masses for the companion as a function of the inclination
angle. Masses of the companion close to that of the primary (for very
low inclination angles) are excluded based on the absence of any
secondary spectra in echelle data. However, the constraint is rather
weak since the S/N of the spectra itself is not very high, making it
difficult to exclude fainter M star companions.

\subsection{Transit Ephemeris}
\label{sec:ephem}

\begin{figure}
  \includegraphics[angle=270,width=8.2cm]{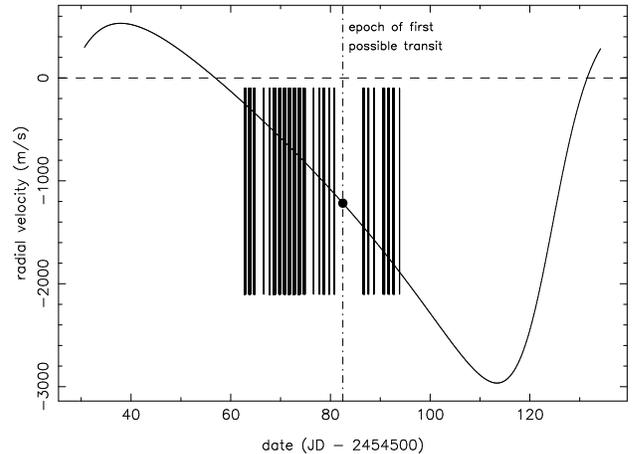}
  \caption{Predicted transit time for the stellar companion. The
    vertical solid lines indicate the dates of photometric coverage of
    the target relative to the transit window.}
  \label{fig:ephemeris}
\end{figure}

The possible transit of the stellar companion motivated a considerable
amount of the photometric campaign that was undertaken (see \S
\ref{sec:phot}). As the knowledge regarding the orbital parametrs
evolved, new transit ephemerides were calculated and transit windows
were communicated to the observers. The time of predicted transit
based upon the complete orbital solution is shown in Figure
\ref{fig:ephemeris}. The solid vertical lines correspond to nights on
which photometric data was acquired. Unfortunately, the final radial
velocity measurements obtained shifted the transit window such that
photometry was not obtained on the night of predicted transit. In
addition, the final orbital solution was checked with the original
SuperWASP observations and did not match any of the observed
transits. Although 6 transits were observed by SuperWASP, they were
only observed in one camera and were not detected by the other cameras
looking at the same field. The photometry presented in \S
\ref{sec:phot} is more than adequate, both in coverage and precision,
to detect the SuperWASP transit if it is real. Thus the SuperWASP
detection is likely to be spurious in nature.

The orbital orientation of this system is such that the periastron
location is behind the star. As such, the probability of a transit
occurring is 0.89\%, but the probability of a secondary eclipse is
1.89\%. These are small odds indeed, but the possibility of
constraining the inclination and therefore the mass of the companion
make this a compelling target to study further. The duration of a
planetary transit for a circular orbit is generally given by
\begin{equation}
  t_{\mathrm{circ}} = \frac{P}{\pi} \arcsin \left(
  \frac{\sqrt{(R_p+R_\star)^2 - (a \cos i)^2}}{a} \right)
  \label{tran_dur1}
\end{equation}
where $R_p$ and $R_\star$ are the radii of the planet and parent star
respectively. For our target, and assuming a companion radius of 1
Jupiter radius, this amounts to a 9.38 hour duration. However, the
eccentricity of the companion changes this duration significantly.
Using the transit duration scaling factor for eccentric orbits
provided by \citet{bur08},
\begin{equation}
  \frac{t_{\mathrm{ecc}}}{t_{\mathrm{circ}}} = \frac{\sqrt{(1 -
      e^2)}}{(1 + e \cos (\omega - \pi/2))}
  \label{tran_dur2}
\end{equation}
we compute a transit duration of 13.61 hours and a secondary eclipse
duration of 6.35 hours. If we assume a radius of 1 Jupiter radius, the
predicted transit depth is 1\%. If we assume a radius of 2 Jupiter
radii, the predicted transit depth becomes 4.2\%. Following up the
target at the predicted transit times is therefore a feasible task
with high scientific yield.


\section{Discussion}

Based on random inclinations the probability of the companion being a
brown-dwarf is high. However, given that low-mass stars in binary
systems are more frequent than brown dwarfs at these separations, it
is necessary to constrain the mass of the companion and exclude the
possibility that it may be a low-mass star. In this section we explore
the various techniques possible to accomplish this and their relevance
and applicability for TYC 2534-698-1

{\it Astrometric Orbit:} Using the orbital parameters, we can estimate
the astrometric perturbation expected for different values of the
companion mass. The orbital parameters yield a semi-major axis of
$0.4417$ AU for the orbit and the best fit spectroscopic stellar model
indicates that the absolute magnitude $M_V$ is 4.77, yielding an
approximate distance estimate of $\sim 155$ parsecs. For the case of
the minimum mass companion, the astrometric perturbation is only 100
$\mu$ arcseconds, and is 220 $\mu$ arcseconds if the companion were
0.08 $M_\odot$. The small semi-major axis coupled with large distance
to the star makes astrometric detection difficult. The Hubble Space
Telescope Fine Guidance Sensors are capable of sub milli-arcsecond
astrometry, but a 1 milli-arcsecond signature only appears when the
companion mass exceeds 0.36 $M_\odot$. Constraining the astrometric
signature to demonstrate that the object is indeed a brown dwarf will
require the astrometric accuracy of planned instruments like Space
Interferometry Mission (SIM).

{\it Transit \& Secondary Eclipse:} The detection of a transit would
constrain the inclination of the companion orbit and unambiguously
confirm its mass to be in the brown dwarf regime. The long period
eccentric orbit however leads to a low probability of only $\sim
0.9$\% for the transit. The predicted transit duration of 13.6 hours
is also too long for the transit to be observed with most available
telescopes. Longer period transits are also more difficult to detect
from the ground due to correlated red-noise and other systematics. The
observed eccentricity and orientation of the orbit makes the detection
of a secondary eclipse more likely than that of a transit. The
probability of this is $\sim 1.9$\% which, while still small, is a
factor of 2 larger than the probability of a primary transit. The
secondary eclipse, if present, would be easier to detect with Spitzer
than similar transits of planets due to the self-luminous nature of
the brown dwarf candidate.

{\it Constraints from Radial Velocity:} High precision radial velocity
observations can themselves be used to exclude certain mass ranges of
the companion. If an iodine cell were used then the presence of a
brown dwarf companion spectra would manifest itself as increased noise
in the radial velocity data when the different spectral chunks are
being compared. Such a technique has been used by \citet{kur08} to
detect a probable brown dwarf around GJ~1046. Meaningful contraints
with this technique however require very high-precision velocities at
2--3 ms$^{-1}$. Even with such a precision it is not possible to
distinguish between a brown dwarf and a very low-mass star. The
primary is also relatively faint, making the acquisition of precision
radial velocities very time consuming.

{\it Interferometric Observations:} The projected maximum separation
of the companion from the primary is only $\sim 2$ mas, making any
direct detection difficult. The projected separation is smaller than
the 5 mas resolution (at 2.2 $\mu$m) of even the 85m Keck
interferometer baseline, and the contrast requirements too high unless
the companion were an early M dwarf at a high inclination.


\section{Conclusions}

We have reported the discovery of a low mass candidate around TYC
2534-698-1. Confirmation of a mass less than $0.08 M_\odot$, or the
detection of a transit, would make this one of the very few known
brown dwarfs in the 'desert'. While the detection of a transit is a
low probability, such a detection would make this the only bright
transiting brown dwarf. The long transit timescale, while
disadvantageous for the detection, is actually a tremendous advantage
for transmission spectroscopy (if a transit is detected) to probe the
atmosphere of the candidate \citep{red08}. In the event that transits
do not occur, tighter limits on the mass of the companion will have to
await future missions such as SIM, which can easily detect the small
predicted amplitude of the astrometric signal.


\section*{Acknowledgements}

The authors would like to thank Leslie Hebb for several useful
discussions, the HET staff astronomers for performing the
observations, and Bertrand Plez for providing the new version of the
turbospectrum code which uses the latest MARCS models. We would also
like to thank University of Texas at Austin for awarding Director's
Discretionary Time which allowed us to obtain complete phase coverage
of the target. The HET is a joint project of the University of Texas
at Austin, the Pennsylvania State University, Stanford University,
Ludwig-Maximilians-Universit\"at M\"unchen, and
Georg-August-Universit\"at G\"ottingen. The HET is named in honor of
its principal benefactors, William P. Hobby and Robert E. Eberly.


\end{document}